# Mitigation Techniques for Cyber Attacks: A Systematic Mapping Study

**Kashif Ishaq**[1] · **Sidra Fareed**[2] ·

[1]School of Systems and Technology, University of the Management and Technology, Lahore, Pakistan

**Corresponding Author:** kashif.ishaq@umt.edu.pk

**Abstract** In the wake of the arrival of digital media, the Internet, the web, and online social media, a flood of new cyber security research questions have emerged. There is a lot of money being lost around the world because of cyber-attacks. As a result, cyber security has emerged as one of the world's most complex and pressing issues. Cyber security experts from both industry and academia institutions are now analyzing current cyber-attacks occurring around the world and developing various strategies to defend systems from possible cyber-threats and attacks. This paper examines recent cyber security attacks as well as the financial losses incurred as a result of the growing number of cyber-attacks. Our findings indicate that the majority of the research chosen for this study focused solely on a small number of widespread security flaws, such as malware, phishing, and denial-of-service attacks. A total of over 50 major studies that have been published in reputable academic journals and conferences have been chosen for additional examination. A taxonomy of cyber-attacks elements that is based on the context of use in various environments has also been suggested, in addition to a review of the most recent studies on countermeasures for cyber-attacks being the state of the art. Lastly, the research gaps in terms of open issues have been described in order to offer potential future directions for the researchers working in the field of cyber security.

## I. Introduction

People are depending on computers and digital technologies more and more. Every day, people, companies, and government organizations use online and offline technology to store and use vital and confidential data. Numerous of these data, including those containing financial and personal information, serve as invitations for hackers to break into computer systems with the goal of eradicating the data and network architecture or extracting information from the data. In the past, a variety of cyberattacks, including email bombing, information or data theft, Denial of Service (DoS) attacks, Trojan attacks, and data or system hacking, have been documented in the literature and online [1,2]. The frequency of attacks is rising daily as a result of the rise of social media, their expanding use, and the majority of people spending the majority of their time online.

Cybercrime refers to the various sorts of attacks carried out over the internet. Cybercrime isn't a new phrase or concept. In 1820, a textile producer in France named Joseph Marie Jacquard created the first loom. This apparatus controlled the weaving of a number of specific fabrics by repeating a series of processes. Employees at Jacquard felt fearful of the process, believing that their traditional jobs and livelihoods were at risk. Employees committed acts of sabotage in order to prohibit Jacquard from employing new technologies in the future. This is known as the first cybercrime to be reported [3-5].

Cybersecurity is the term used to describe the safeguarding of data stored on client PCs, servers, and internet-based data transport. Attacks on communication networks using the internet the widespread use of social media and information technology, as well as web-based Internet computing and related services like cloud computing, mobile commerce, and health informatics, have heightened the demand for cyber security. One of the biggest and most urgent issues the world has ever seen has developed. As a result, in order to create a security strategy for protecting information and communication infrastructures that are vulnerable to cyberattacks, it is essential to comprehend both recent and historical cyberattacks as well as mitigation strategies.

The goal of this paper is to perform a methodical mapping analysis in order to find and evaluate common cyber security issues. Therefore, this paper examines recent cyber-attacks, their results, and available defense mechanisms. The paper examines current and recent cyber-attacks, the implications of cyber-attacks are examined, as well as mitigation strategies. The novelty of this work is the taxonomy of cyber security elements presented to help the researchers and mitigation techniques with their frequency and percentage is also briefly described. Lastly, research gaps in terms of open issues have been described in order to offer potential future directions for the researchers.

The rest of this paper is organized as follows. The related work is briefly described in Section 2. In Section 3, the research technique is described. Section 4 presents the study's findings, and Section 5 presents a commentary of those findings. Section 6 discusses some unresolved issues after that. Section 7 concludes the paper.

## II. Related Work

The systematic literature reviews (SLR) and mapping studies on the topic of the cyber environment are covered in the following areas.

Hydara et al. [2] analyzed web apps for cross-site scripting (XSS) vulnerabilities using an SLR. Despite the fact that they discovered a number of strategies in their research to address XSS vulnerabilities, the researchers claim that there is presently no one solution that can be used to minimize the problem of XSS.The results of the SLR suggest that more research is needed to address the problem of removing XSS from source code before deployment. A thorough mapping study on the security of cyber-physical systems was carried out by Lun et al. [3]. Among the domains looked at were network systems, smart grids, information systems, and autonomous controls. The study's conclusions show that researchers have just focused on smart grid systems and placed a special emphasis on physical-level threat.

Mishna et al. [4] also wrote an SLR in order to evaluate the existing state of the art with regard to techniques that can be used to protect and restrict child cyber-abuse. The study's goal is to assess how well cyber-abuse therapies work in promoting Internet safety awareness and reducing risky online behavior. The findings demonstrate that while cyber abuse intervention can raise public awareness of safety issues, there is no concrete link between it and risky online conduct.

In order to look into the procedures for evaluating cyber security awareness, Rahim et al. [5] used an SLR. The study's findings indicate that a number of methods for increasing cyber security awareness have been proposed in the literature. To achieve better results, it is vital to employ a range of techniques. As young people are the primary targets of cyberattacks, there is also a need to raise knowledge of cyber security.

An SLR on cyber situational awareness was conducted by Franke and Brynielsson [43], and it was based on 102 main publications that had been published up until 2013. According to their findings, many aspects of online situational awareness have been researched and developed to varying degrees. Some of these aspects are more advanced than others. In contrast to Franke and Brynelsson's research, which focused on cyber situational awareness, our investigation's primary objective was to determine which cyber security threats and vulnerabilities are most widespread. In contrast to Franke and Brynielsson's exhaustive study of the existing literature, we opted to carry out research in the form of a systematic mapping. In addition, items that are relevant to our inquiry were still being published right up until the year 2018 came to a close.

In order to establish the current state of the art about the various infrastructures that now allow cyber foraging, Lewis and Lago [6] carried out an SLR study. It's called "cyber foraging" when low-powered computers outsource their heavy lifting to nearby, more powerful ones. Many architectural options for transmitting data and computation to mobile devices were investigated as part of the research. In order to help academics and professionals in the field of architecture improve their designs to facilitate cyber foraging, the writers analyzed preexisting structures in order to isolate their component parts and then codified those parts as architectural methods. The study by Ramaki et al. [8], who did extensive mapping of intrusion alarm analysis using the SMS approach. In order to address the research topics, 411 studies were analyzed as part of this mapping project. Analysis of intrusion alerts is a burgeoning field of study, according to the study's findings the report summarizes the most current advances in intrusion warning analysis at the time but no previous analysis has been discussed.

Using the Global System for Mobile Communications (GSMC), Enoch et al. [7] attempted to document possible attack scenarios for dynamic networks (GSM). A change in network properties is used to analyze the impact of a security metric adjustment. As a result of current security concerns, each statistic was examined. Helps network administrators and practitioners and academics determine the best security metrics from this information. Rather than focusing on a single cyber security danger, this study focused on a wide range of security flaws.

Researchers evaluated a Bayesian network model's contribution to the field of cyber security by referring to Chockalingam et al. [9]'s in-depth review. Within the scope of this investigation, 17 different Bayesian network models were found and analyzed. Bayesian network models can be used to solve the issue of malicious insiders, according to the findings of the study. Industrial control systems, on the other hand, are more commonly employed to handle security concerns in the IT environment than these models. There are not any general-purpose Bayesian network models, either, that can deal with all the many kinds of cyber security issues. Alguliyev et al. [10] conducted a study to identify current attacks and existing defenses, and analyzed the literature on SCADA and smart grid security. The paper's main contributions are the analysis of cyber-attack tactics, modelling of the effects of these attacks, and creation of security architecture.

The fact that cyber security and cyber awareness have been the topic of substantial research is made abundantly evident by the discussion that was just presented. SLRs as well as investigations using a systematic mapping approach have been finished. However, the majority of the mapping studies that are currently being conducted are concentrating their attention on the protection of cyberspace and, in particular, the promotion of cyber awareness. Studies do not systematically map out the most critical cyber security flaws or the approaches that can be used to reduce the dangers associated with these flaws. This mapping study's objective is to contribute researchers with an understanding of the vulnerabilities that are currently present in cyber security, as well as the strategies that can be implemented to detect and mitigate the holes that have been identified in order to fill the informational void that has been identified.

The paper examines current and recent cyber-attacks. The implications of cyber-attacks are examined, as well as mitigation strategies. The novelty of this work is the taxonomy of cyber security elements presented to help the researchers.

## III. Research Methodology

The recommendations for carrying out a systematic mapping investigation were followed in this work [16]. This approach is preferred for a variety of reasons. It is a methodical and organized strategy to find, assess, and analyze all the studies that are pertinent to a specific research question, area of focus, or criteria of interest. Finding the gaps and missing data in the recent research and giving academics or industry professionals the background data, they need to support new research are all goals of a systematic mapping study, which refers to an approach that is both well-defined and disciplined, and may be used to analyze and compile the empirical data that is associated with a method or technology. While a systematic mapping study requires more time and effort than a traditional literature review, it offers a more understanding of the subject and a solid foundation for making statements regarding research issues [17]. Four distinct steps make up a systematic mapping study methodology, as seen in Fig. 1.

The specifics of each step taken in the current investigation are detailed in a protocol for a SLR that has been created. The following is a basic outline of the main steps:

1. Coming up with research question ideas.

2. The search strategy as well as the search keyword need to be defined.

3. Providing details on the criteria for participation in the study as well as those for being excluded from it.

4. Extracting data and aligning it with the data to answer the research questions.

5. Results extraction and data analysis.

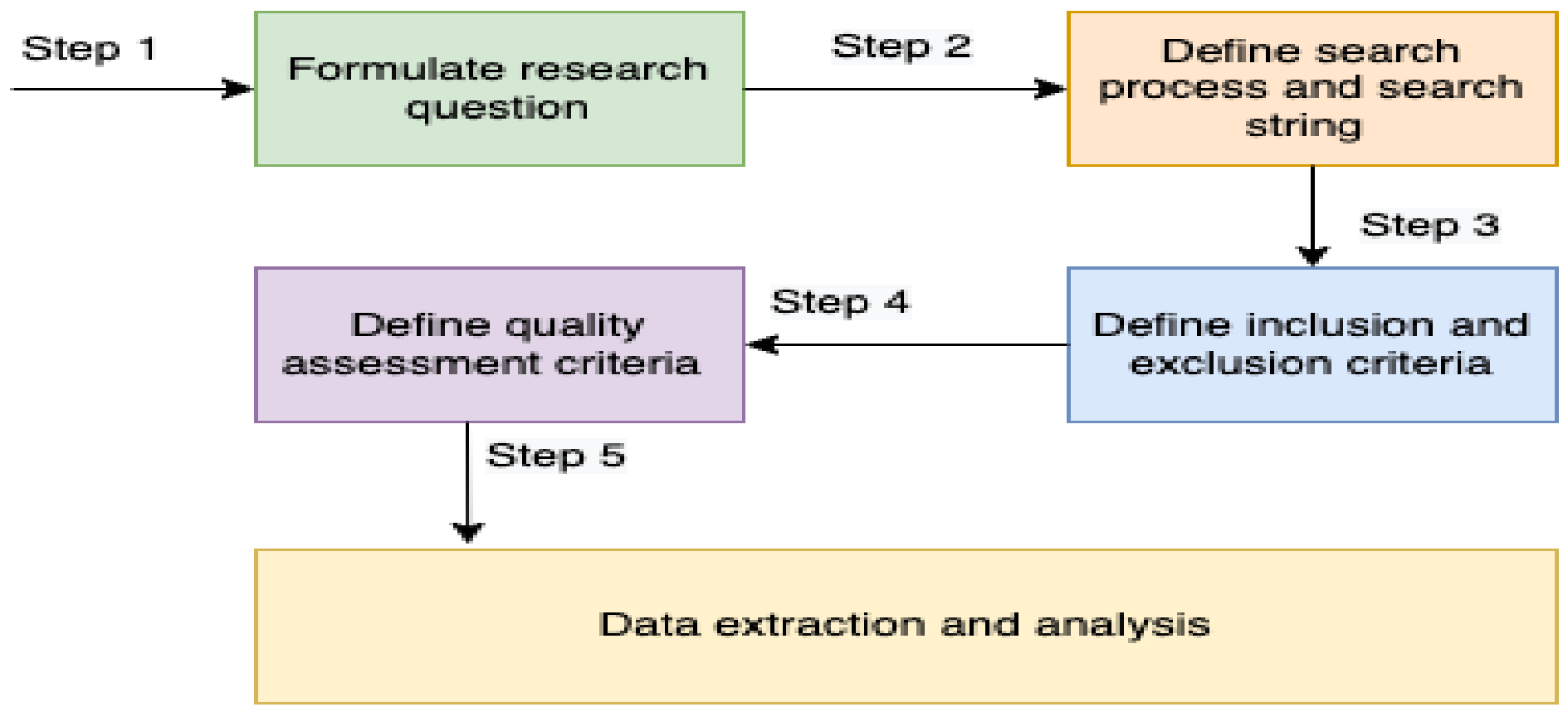

FIGURE 1. **Stages of a systematic mapping study**

### A. RESEARCH OBJECTIVE

The prime intents of this research are as follows:

RO1: The majority of the common vulnerabilities in cyber security have been found and investigated.

RO2: The purpose of this mapping study is to discover studies on cyber security vulnerabilities that are now available and to categories the remedies for such vulnerabilities.

RO3: Find major victims of cyber threat.

RO4: Identify methods of data collection and mitigation approaches.

RO5: Identification of the main issues and unsolved challenges.

### *B. RESEARCH QUESTIONS*

The primary research questions have been identified in order to efficiently conduct out this SLR initially. Additionally, a thorough search strategy that is necessary for the review's identification and extraction of the most important articles has been devised. Table I lists the research questions that were the focus of this evaluation along with their primary drivers. The queries are addressed and answered in light of the established method.

**TABLE I. RQ and major motivations**

| | Research Question | Major Motivation |
|---|---|---|
| **RQ1** | Which are the publications channels on cyber security threats and attacks. | The research question aims to identify the primary outlets for publishing academic works on the subject of cyber security. Determining the solution to this inquiry can assist scholars in identifying the primary conferences and publications within the field. |
| **RQ2** | Who are the main individuals effected by security flaws? | The response to this question emphasis the people who have been harmed by security breaches. Individual and organizational victims are the two basic categories in which we classify victims. |
| **RQ3** | In the selected studies, which apps are the targets of cybercrimes? | The response to this query will include a list of the applications that were the subject of the research that were chosen, together with data about the users of these applications so that they may be protected against cyberattacks. |
| **RQ4** | What frequent methods of cyber security mitigation are covered in the literature? | Researchers will be able to have a better understanding of currently accessible mitigation options after receiving an answer to this question. |
| **RQ5** | What are some typical threats and flaws in cyber security? | Finding the security flaws that are most frequently present in the chosen research is one of the most crucial RQs. Finding a solution to this issue will help in determining the most significant security weaknesses and the most crucial research directions in the area. |

### C. *SEARCH SCHEME*

Before formally beginning the mapping work, ScienceDirect was searched for "empirical studies on cyber security." ScienceDirect was chosen since it is a well-known library with a sizable number of papers from numerous fields. This preliminary search was carried out with the intention of accomplishing two tasks: first, determining whether there are sufficient studies to carry out an SLR, and second, locating some primary research that might potentially be used in the future to validate the search keyword. The chosen studies were imported into Endnote [48] via export. After reviewing the abstracts of the papers that had been acquired, 9 empirical studies were chosen to serve as the primary research that would be used to the validation of our modified string. This research would be applied to the validation of our changed string. Numerous empirical studies were discovered during this informal search process; thus, it was decided to do a systematic mapping research.

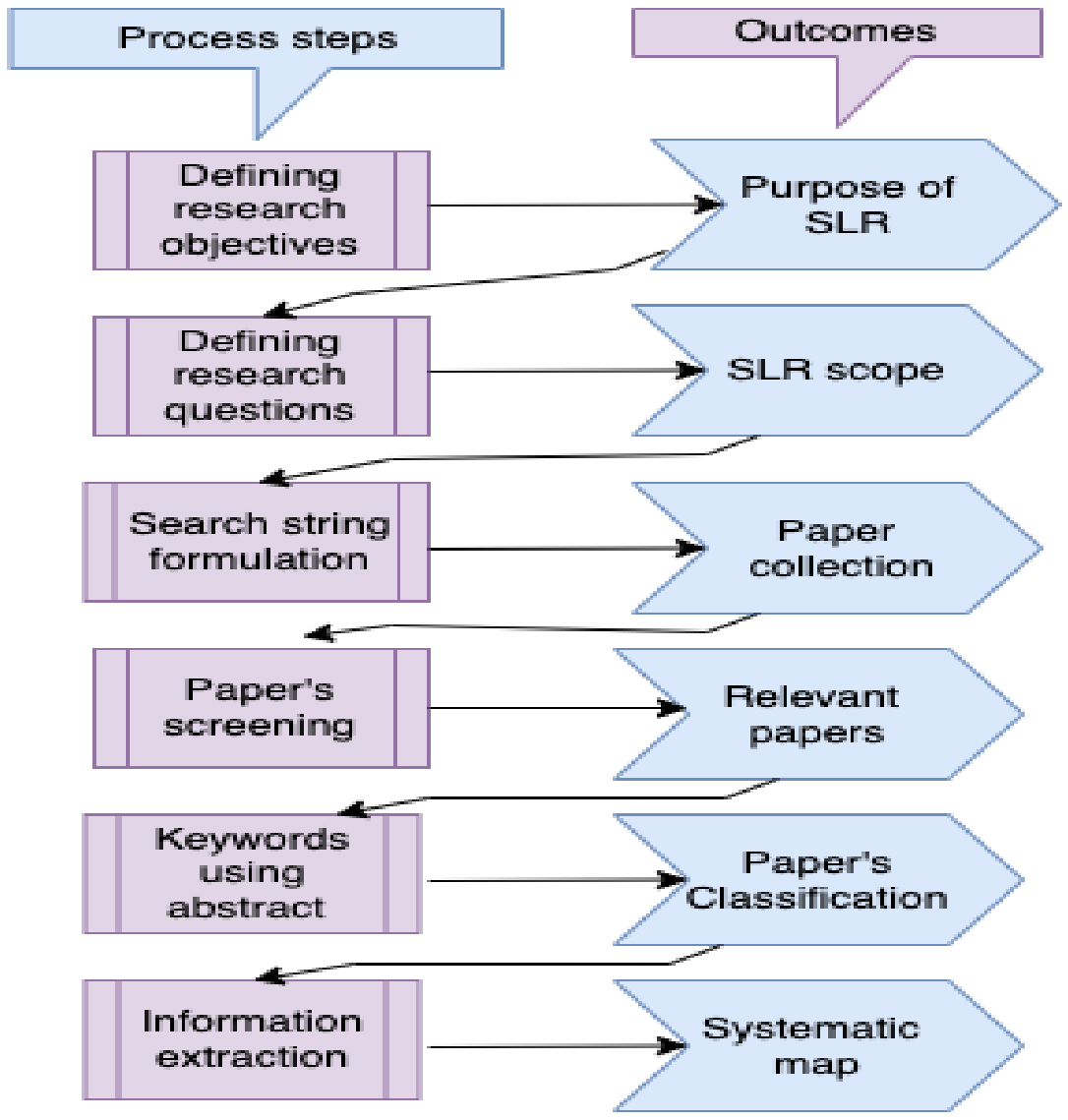

**FIGURE 2.** **An SLR process mode**

Cyber AND security was the first string defined for the search procedure. When this preliminary string was submitted into the search engine for ScienceDirect, the findings that were returned by the engine did not include all of the primary research. In addition, the search phrase was analyzed, and recommendations from two senior academic software engineering experts who have worked on SLRs were solicited. These experts' recommendations led to a revision of the original string and a division of the main search string into two pieces. Information can be swiftly evaluated and validated using expert opinion [49]. Here are the two components of the string we defined.

1. Cyber security.
2. Attack/threat/vulnerability.

There are three parts to the whole search string. There are three parts of the string that can be used to limit the results to those that are computer-based or computational, peptide prediction-related, and studies based on methods other than computation. The first part of the string can be used to limit results to those that are computational (i.e. Lab-based experiments). An equation for the search string is Equation (1), which is a mathematical representation of the search.

$$\boldsymbol{R} = \forall\ [(\boldsymbol{CS}\ \vee \boldsymbol{C}\mathbf{P}\ \vee \boldsymbol{IS}) \wedge (\boldsymbol{NS}\ \vee \boldsymbol{ITS}\ \vee \boldsymbol{CS}) \not\equiv (\boldsymbol{SS}\ \vee \boldsymbol{CT}\ \vee\ \boldsymbol{C}\ \vee\ \boldsymbol{V} \vee \mathbf{R})]$$

Here in (1), R refers to the search results that were obtained against the search string, represents for all, '∨' is used for the 'OR' operator, and '∧' is used for the 'AND' operator. Together, these symbols and the search phrases expressed in Table 2 formalize the entire search string according to each chosen repository.

To compile all potential pertinent research, the synonyms of both of these sections of the string were taken into consideration. Since the results of the validation against the list of primary studies were encouraging, it was decided to use this revised string going forward for data extraction. The fact that the results of the second search string include all of the primary studies that were selected for additional investigation is evidence that the first search string is accurate. The following three crucial steps serve as the foundation for the search method in this systematic mapping study:

### 1) SEARCH STRING

After going through a number of different iterations and revisions, for this mapping experiment, the following search string was chosen as the best option in the end:

Cyber OR Privacy OR {cyber physical} OR {cyber security} OR {IT security} OR {Network security} OR {Internet security} OR {Software Security} OR {Computer Security}) AND ({cyber threat} OR (vulnerability OR {cyber-Attack} OR {cyber Terrorism} {OR violence OR challenge OR risks).

The following online libraries conducted the search using the final search string, which was additionally modified to fit the search functionality each library offered:

IEEE Explore; ScienceDirect; ACM Digital Library; SpringerLink; John Wiley Online Library, publisher wise search string is in table 3.

The aforementioned five databases were chosen because they are well-liked places to post research on cyber security. These datasets have also been used by other researchers to conduct SLR studies [35, 50, 51].

The most recognized and topic-specific journals have been picked to conduct the literature search from online repositories, dedicated to research publication and gathering. Table 2 lists the specifics of the chosen repositories, the search terms used, and the outcomes.

**TABLE II. Terms and keywords used in search**

| Terms (Keywords) | Synonyms / Alternate Keywords |
|---|---|
| **+ Cyber Security (CS)** | Cyber(C), Privacy(p) |
| **+ Cyber Attack (CA)** | Malware, hacks |
| **- Cyber threat (CT)** | Cyber attack |
| **- Risks (R)** | - |

**TABLE III. Publisher wise search strings**

| *Repository* | *Search Strings* |
|---|---|
| *John Wiley Online Library* | *("CYBER" OR CYBER SECURITY OR PRIVACY AND VULNERABILITY OR CYBER THREAT OR CYBER CRIME)* |
| *Springer Link* | *(("CYBER" OR CYBER SECURITY OR PRIVACY) AND (VULNERABILITY OR CYBER THREAT OR MITIGATION OF THREATS) AND NOT (PHYSICAL SECURITY))* |
| *Science Direct* | *((CYBER SECURITY OR PRIVACY OR CYBER) AND ("CYBER THREAT OF COMPUTER" OR "VULNERABILITY" OR "SOFTWARE ") NOT (PHYSICAL SECURITY))* |
| *IEEE Xplore* | *(((((("ALL METADATA":"SECURITY BASED") OR "ALL METADATA": CYBER PROTECTION) OR "ALL METADATA": IT SECURITY) AND "ALL METADATA": CYBER THREAT) OR "ALL METADATA": VULNERABILITIES)* |
| *ACM Digital Library* | *(((("CYBER" OR "PRIVACY") AND BASED) OR CYBER SECURITY OR (("PHYSICAL" AND " CYBER ") OR "IT SECURITY")) AND ((("CYBER THREAT" OR) "VULNERABILITY" )AND ("THREATS" OR "THREAT") AND CYBER-ATTACK) OR ((" VIOLENCE " OR "RISKS") AND CHALLENGE) )))* |

**TABLE IV. Study Selection (inclusion exclusion)**

| Source | Final selection | Retrieved |
|---|---|---|
| **Wiley** | 4 | 91 |
| **SpringerLink** | 8 | 1445 |
| **ScienceDirect** | 18 | 1308 |
| **ACM** | 10 | 323 |
| **IEEE** | 10 | 3897 |
| **Total** | 50 | 7064 |

### *D. INCLUSION AND EXCLUSION CRITERIA*

This section emphasizes the method used to choose pertinent publications in accordance with the research questions and describes the criteria for inclusion and exclusion that were utilized during the publication selection process.
Parameters defined for inclusion criteria (IC) are:

IC 1) empirical research with a primary focus on vulnerabilities in computer and network security.
IC 2) Studies that are aimed at finding a solution to cyber security flaws.
Such research including laboratory-based experimental identification / prediction of peptides for their therapeutic benefits, or other inappropriate approaches, were excluded from the review process.
EC 1) studies that don't offer clear guidance on how to find cyber security problems.
EC 2) Papers that do not focus on cyber security.
EC 3) studies whose results are not practically tested.
EC 4) Research where the conclusions are not subjected to an empirical evaluation.

The process of choosing publications was automatic and consisted of two steps: first, the search results were initially chosen based on the selection criteria by screening the publications' titles and abstracts; second, in order to construct a shortlist of publications from which one would be chosen based on the inclusion criteria, the articles that were picked in the original round were given thorough consideration and examination. Table 4 shows the total number of papers included in this paper.

# IV. Results

1) RQ1: Which are the publications channels on cyber security threats and attacks?

This question of this study examines the chosen publication's venue and source type, with the intention of addressing research question 1 (RQ1), which pertains to the primary venues of publication that make significant contributions to the field of cyber security. In order to conduct an analysis on venue and source type, we have identified five libraries as the primary venues for publishing, as indicated in Tables 4 and 5. The selected studies obtained from these libraries were disseminated through three principal channels of academic publication, namely conferences, journals, and workshops. The distribution of papers throughout conferences and publications demonstrates a close balance, with only three studies being published in workshops. The distribution of scholarly research publications across conferences, journals, and workshops was found to be 48%, 48%, and 4%, respectively. The findings shown in Table 4 indicate that the libraries of IEEE and ACM encompass a greater number of conference papers in comparison to journal papers. Out of the publications obtained from the IEEE collection, only three were found to be published in IEEE journals, while the remaining papers were found to be published at IEEE conferences. at the context of the ACM collection, it was observed that the papers collected for analysis were predominantly published at conferences and workshops, accounting for 67% and 33% respectively. Notably, no journal papers were included in the dataset relevant to the specific subject being investigated. Table 5 illustrates the prominent locations for the key inquiries, each of which has a frequency of 2 or more. The results indicate that a significant proportion of the articles are disseminated through academic journals. The Journal of Computer & Security, which is published by ScienceDirect, exhibits the most prolific publication rate of articles, accounting for 5 out of the total 50 articles. Following closely is Information Sciences, also published by ScienceDirect, with 3 out of the 50 articles.

TABLE V. **Summary of publication sources**

| Source | Type | Library | Frequency |
|---|---|---|---|
| **International Journal of IS** | Journal | Springer | 3 |
| **Computer Networks** | Journal | ScienceDirect | 7 |
| **Procedia Technology** | Journal | ScienceDirect | 2 |
| **International Conference on Recent Trends in IT** | Conference | IEEE | 5 |
| **Winter Simulation Conference** | Conference | ACM | 3 |
| **Annual Cyber Security and Information Intelligence Research Workshop** | Workshop | ACM | 2 |
| **Computers & Security** | Journal | ScienceDirect | 5 |
| **Security and Communication Networks** | Journal | Wiley Online | 2 |
| **Future Generation Computer Systems** | Journal | ScienceDirect | 2 |
| **IEEE Access** | Journal | IEEE | 5 |
| **Annual Conference on cyber and information security research** | Conference | ACM | 2 |
| **International Conference on Advanced Communication Technology** | Conference | IEEE | 7 |
| **Information Sciences** | Journal | ScienceDirect | 5 |

2) RQ2: Who are the main individuals effected by security flaws?

The second research question, who are the main victims of these security flaws, is focused on identifying the major victims of cyber security vulnerabilities. Table 6 displays the findings after categorizing the victims into two main groups: organizations and people. In the chosen research, some vulnerabilities had an impact on both people and organizations simultaneously, hence the findings for these vulnerabilities overlap.

**TABLE VI. Victim Frequency**

| Victim | Response %age | Responses (%) |
|---|---|---|
| Organization | 74 | 95 |
| Individual | 9 | 11 |

3) RQ3: In the selected studies, which apps are the targets of cybercrimes?

The third research question's (RQ3) main objective was to identify the apps that the chosen studies' top cybercrime targets. We divided the information about the organizations and applications of the intended victims that we retrieved from the chosen research into the following three groups.

### i. Infrastructure That Was Targeted

Based on the information that was collected, the following types of infrastructure were a primary target of cybercriminals:

- Smart grid.
- Social media.
- Mobile software.

- Control system for use in industrial settings.
- Network.
- A network that is distributed.
- Software hosted on the cloud
- Several VLAN.
- Computer networks and physical applications.
- Application Servers to be Used.
- Peer-to-peer, or P2P, computer networks.
- A network that is created on demand for automobiles (VANET)
- Information systems and the IOT.
- Software with a client-server architecture
- Online data.
- Collaborative working nodes that are connected to one another over an MPLS-VPN cloud.
- A network entry point for commercial enterprises

### ii. Target Applications

According to the statistics that we have obtained, the following applications were the victims of malicious cyberattacks:

- Platform for neuromorphic hardware that uses little energy.
- 24. 8. 0 Thunderbird
- Web-based program.Xen 4.4.0
- E-commerce.
- Hacker database.
- Banking.
- Libav 10.1

Agencies and organizations that were the focus of the attack According to the findings of our investigation, the following companies and organizations were the targets of cyberattacks:

- AhnLab Security Emergency Centre.
- Aircraft attitude sensors.
- DARPA

4) RQ4: What frequent methods of cyber security mitigation are covered in the literature?

Our final research question (RQ4) focused on identifying the cyber threat mitigation strategies employed by distinct victim industries. The frequency and proportion of different mitigation measures that are utilized to safeguard the cyber environment from potential security hazards are outlined in Table 7. Data extraction revealed that several firms used multiple security mitigation techniques to safeguard their online environments. For instance, many cyber organizations employed firewalls and IDs in addition to other security measures. Numerous articles have also employed traffic analysis to identify security attacks. According to our analysis, intrusion detection systems and firewalls are the most often employed methods for preventing cyberattacks (17 out of 50 and 13 out of 50). With a frequency of occurrence of 7 out of 50, traffic analysis was the second most frequently employed method of cyber-attack mitigation. Measures based on signatures and to prevent Antiphishing [8], which occur 6 out of 78 times each, are the third most frequently employed methods of thwarting cyberattacks. Table 7 lists the remaining mitigating strategies and how frequently they occur. Because some articles listed many methods for thwarting cyber security attacks, the overall frequency is higher than 50. Nine of the collected papers, however, omitted the name of the mitigation method that was applied.

**TABLE VII.** **Cyber Attack Mitigation technique**

| Mitigation techniques | Frequency | Percentage (%) |
|---|---|---|
| **Sandboxing** | 3 | 4 |
| **Signature and anomaly-based detection** | 6 | 8 |
| **Improving the way of accepting incoming requests** | 2 | 1 |
| **Anti-malware software** | 5 | 6 |
| **Firewalls** | 13 | 17 |
| **Darknet** | 2 | 3 |
| **Command and control (C&C) servers** | 2 | 1 |
| **Content-based spam fltering technique** | 3 | 4 |
| **Iterative approach of critical component identifcation** | 3 | 4 |
| **Algorithm weakly supervised** | 5 | 6 |
| **Not mentioned** | 9 | 12 |

5) RQ5: What are some typical threats and flaws in cyber security?

The findings of the systematic mapping investigation are presented in this section. 162 studies in all were chosen for the initial search phase. In the end, 50 articles were chosen based on the inclusion and exclusion criteria. Table 4 displays specifics for each iteration, to answer the study questions, these chosen papers were carefully examined.

A thorough mapping investigation (RQ5) uncovered the following key types of cyber security risks and vulnerabilities. Table 8's Column 1 lists the cyber security issues uncovered throughout this mapping assessment. When a vulnerability was found in the selected studies, the frequency of occurrence and the percentage of occurrence is shown in Column 2 and Column 3 of Table 9. A few of the significant vulnerabilities we discovered in our mapping study include: Malware, DOS, SQL injection, Torjanhorse, Phishing, man-in-the-middle, XSS and credential reuse are all examples of security threats. The systematic mapping analysis focused the most on the denial-of-service vulnerability (37 percent). Malware (21%) and phishing (22%) attacks are the next two vulnerabilities that are most commonly mentioned in the literature. Table 8 displays specifics of the remaining vulnerabilities.

TABLE VIII. **Cyber security flaws and vulnerability Categorization**

| Threat and vulnerability | Frequency | Percentage (%) |
|---|---|---|
| **Cross-site scripting (XSS)** | 1 | **1** |
| **Credential reuse** | 1 | **1** |
| **Denial-of-service (DoS)** | 29 | **37** |
| **Phishing** | 7 | **9** |
| **Malware** | 16 | **21** |
| **Session hijacking and man in-the-middle attacks** | 2 | **3** |
| **SQL injection attack** | 3 | **4** |
| **Other** | **19** | **24** |

### *E. RESEARCH AND PRACTICAL IMPLICATIONS*

There are both theoretical and practical applications to this mapping work. We grouped the major security flaws and determined how frequently they appeared in the chosen research. This will enable researchers to determine which security flaws require further study. Future research can focus on the security vulnerabilities that require additional study. In addition, we divided the studies into groups based on where they were published. This will aid in the analysis of the sociocultural influences on cyber security.

It is also hoped that the identification of the primary vulnerabilities and the frequency of their occurrence would aid practitioners in creating campaigns to inform people and organizations about these vulnerabilities and their countermeasures. It is customary to draw attention to cyberattacks that happen frequently because not all assaults and vulnerabilities are created equal. Additionally, it will direct financial choices in crucial security domains. With the use of this SLR and the empirical findings revealed in this work, it will be possible for practitioners to make investment decisions while simultaneously developing tools and methods to defend the cyber environment. This will be possible because of the work that was done.

Cyber firms must give their customers instructions and training regarding important vulnerabilities and how to stay safe. Organizations should create methods to draw up appropriate privacy rules to safeguard both individuals' and organizations' valuable assets. In order for the client to use attack detection tactics and tools with ease, organizations should carefully choose them. Additionally, businesses must ensure that staff members do not share their personal information with any outside parties or reply to spam emails or text messages.

### *F. QUALITY ASSESSMENT*

Conducting a comprehensive evaluation of the chosen articles is a crucial undertaking within the framework of a systematic review. Due to variations in study design, the evaluation of quality has been conducted through the utilization of a sequential assessment procedure. The internal criteria are used to evaluate an article's internal quality, while the stability and dependability of the publication source are taken into account to establish an article's external quality. The "Computer Science Conference rankings (CORE)" and the "Journal Citation Reports (JCR)" have been used to rate and assess the external quality. The individual scores for each category are added to create the overall score. The final score can range from a minimum of 0 to a maximum of 10, and it can be classified as high ranked if it is larger than 8, ordinary ranked if it is between 6 and 8, and low ranked if it is less than 6.

TABLE IX. **Quality Assessment Score**

| Ref. No. | P.Type | Year | Research Approach | Domain | I-Score | E-Score | Total Score |
|---|---|---|---|---|---|---|---|
| [1] | Conference | 2017 | SLR | ACM | 7.5 | 2 | 6 |
| [2] | Conference | 2016 | SLR | ACM | 4.5 | 2 | 6 |
| [3] | Workshop | 2013 | SLR | ACM | 8 | 2 | 10 |
| [4] | Workshop | 2008 | SLR | ACM | 7 | 2 | 9 |
| [5] | Conference | 2017 | SLR | ACM | 6.5 | 1 | 8 |
| [6] | Conference | 2011 | SLR | Wiley | 5.5 | 2 | 7 |
| [7] | Conference | 2017 | SLR | Wiley | 5 | 2 | 7 |
| [8] | Workshop | 2017 | SLR | Wiley | 6.5 | 2 | 8 |
| [9] | Conference | 2007 | SLR | Wiley | 4 | 2 | 6 |
| [10] | Conference | 2015 | SLR | Springer | 5 | 2 | 7 |
| [11] | Journal | 2009 | SLR | IEEE | 5 | 2 | 7 |
| [12] | Conference | 2017 | SLR | IEEE | 7 | 1 | 9 |
| [13] | Conference | 2014 | SLR | Springer | 7 | 2 | 9 |
| [14] | Journal | 2018 | SLR | IEEE | 6 | 2 | 8 |
| [15] | Conference | 2017 | SLR | Springer | 6 | 2 | 8 |
| [16] | Journal | 2017 | SLR | IEEE | 6 | 2 | 8 |
| [17] | Conference | 2014 | SLR | Springer | 6 | 1 | 7 |
| [18] | Conference | 2015 | SLR | IEEE | 8 | 1 | 9 |
| [19] | Conference | 2013 | SLR | IEEE | 4.5 | 1. | 6 |
| [20] | Conference | 2012 | SLR | IEEE | 4.5 | 2 | 6 |
| [21] | Conference | 2014 | SLR | IEEE | 6 | 2 | 8 |
| [22] | symposium | 2013 | SLR | IEEE | 5.5 | 2 | 7 |
| [23] | Conference | 2015 | SLR | IEEE | 7 | 2 | 9 |
| [24] | Conference | 2014 | SLR | IEEE | 6 | 2 | 8 |
| [25] | Journal | 2014 | SLR | IEEE | 5.5 | 2 | 7.5 |
| [26] | Conference | 2016 | Evaluation | IEEE | 7.5 | 2 | 6 |
| [27] | Conference | 2015 | SLR | IEEE | 4.5 | 2 | 6 |
| [28] | Conference | 2015 | SLR | IEEE | 8 | 2 | 10 |
| [29] | Conference | 2016 | SLR | IEEE | 7 | 2 | 9 |
| [30] | Conference | 2009 | SLR | IEEE | 6.5 | 1 | 8 |
| [31] | Conference | 2017 | SLR | Springer | 5.5 | 2 | 7 |
| [32] | Conference | 2014 | SLR | IEEE | 5 | 2 | 7 |
| [33] | Conference | 2016 | SLR | Springer | 6.5 | 2 | 8 |
| [34] | Conference | 2011 | SLR | Science Direct | 4 | 2 | 6 |
| [35] | Conference | 2012 | SLR | Science Direct | 5 | 2 | 7 |
| [36] | Conference | 2016 | SLR | Springer | 5 | 2 | 7 |
| [37] | Conference | 2017 | SLR | IEEE | 7 | 1 | 9 |
| [38] | Conference | 2010 | SLR | Springer | 7 | 2 | 9 |
| [39] | Journal | 2017 | SLR | IEEE | 6 | 2 | 8 |
| [40] | Conference | 2012 | SLR | Science Direct | 6 | 2 | 8 |
| [41] | Conference | 2018 | SLR | IEEE | 6 | 2 | 8 |
| [42] | Journal | 2019 | SLR | IEEE | 6 | 1 | 7 |
| [43] | Journal | 2013 | SLR | Science Direct | 8 | 1 | 9 |
| [44] | Journal | 2015 | SLR | Science Direct | 4.5 | 1. | 6 |
| [45] | Journal | 2012 | SLR | Science Direct | 4.5 | 2 | 6 |
| [46] | Journal | 2016 | SLR | Science Direct | 6 | 2 | 8 |
| [47] | Journal | 2015 | Solution proposal | Science Direct | 5.5 | 2 | 7 |
| [48] | Journal | 2017 | Solution proposal | Science Direct | 7 | 2 | 9 |
| [49] | Journal | 2018 | Evaluation | Science Direct | 6 | 2 | 8 |
| [50] | Journal | 2018 | Solution proposal | Science Direct | 5.5 | 2 | 7.5 |

## V. Gaps and Challenges

Cybersecurity is a rapidly increasing field of research because of its pervasive application in nearly every facet of daily life. On the other hand, it places stringent requirements on the integrity and protection of computer networks against assaults from within as well as from without. To address the major security risks in this area efficiently, fundamental research is needed. In order to help researchers, identify gaps in the body of existing knowledge and new areas for investigation, we have identified significant and often occurring cyber security vulnerabilities in this study. The following are some potential future research areas:

The common cyber security flaws are shown in Table 8 along with a frequency of occurrence. This means that malware and denial-of-service attacks are frequent causes of security vulnerabilities. It is essential to work on developing strategies that can defend the online environment from the aforementioned vulnerabilities.

Table 6 illustrates the proportion of individuals and institutions that were the focus of the investigation. Even if organizations are more likely than individuals to experience security problems, a trustworthy information security system must nonetheless be developed to protect sensitive data. A safe and transparent framework must be created to protect organizations from both internal and external security attacks. The infrastructure, apps, and businesses that are the main targets of cybercrime are listed in Section 5. This demonstrates the need for mitigation methods to be put out in order to shield these environments from online threats.

## VI. Taxonomy

Five libraries were taken into consideration as the primary venues for publications for the venue and source type analysis, as shown in Table 4. Conferences, journals, and workshops were the three main publication forms in which the chosen papers from these libraries were published. Although there were approximately equal numbers of studies published in journals and conferences, there were only three papers published in workshops. The percentage of studies that were published in workshops, conferences, and journals was 48%, 48%, and 4%, respectively. There are more conference papers than journal publications in the IEEE and ACM libraries. Only three of the papers taken from the IEEE collection were published in periodicals issued by the IEEE; the remaining publications were all from IEEE conferences. In the case of the ACM library, all of the papers that were retrieved (67 % and 33 %, respectively) were presented at conferences or workshops, and no journal articles on the subject of the research were discovered. Figure 3 shows taxonomy of cyber threats.

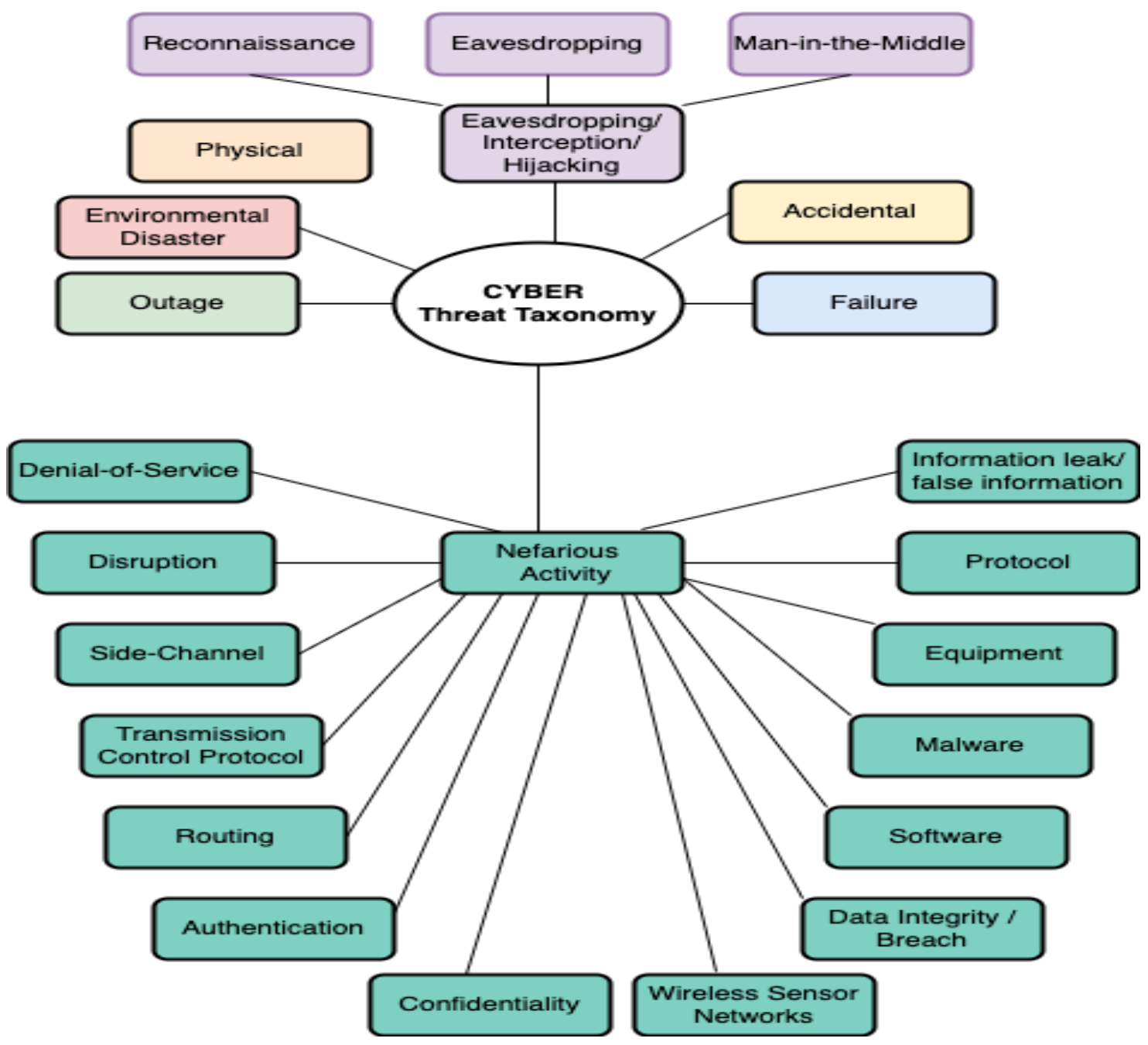

FIGURE 3. **Taxonomy of cyber threat**

## VII. Conclusion

The results of an in-depth mapping analysis that were undertaken to try to identify and evaluate the most widespread cyber security flaws are presented in this research. The investigation was carried out by the authors of this paper. A number of important discoveries were made, including the following:

In total, 134 articles were chosen for this methodical investigation of mapping based on the results of a predetermined search query.50 articles that satisfied our inclusion criteria were chosen after all the papers had been screened. Seven significant security vulnerabilities that were discussed the most in the chosen papers were retrieved after careful analysis of each publication. According to our data, malware and denial-of-service vulnerabilities were often reported security flaws, with a frequency of 37% and 21%, respectively. According to the selected study, algorithm-based solutions, machine learning techniques, and intrusion detection systems are the most popular methods for finding these vulnerabilities. According to our analysis, businesses are more susceptible to cyberattacks than people are. There are, however, some attacks that target both people and organizations. Phishing attacks mostly target individuals, who receive spam emails and instant chats intended to reveal their personal login information. Cyber awareness is required to educate people about cyber-attacks and alert them to the disclosure of their personal information. Cloud computing, smart grids, the IOT, and cyberspace appear to be the most popular places for criminals to operate. These cyber environments must be planned, designed, implemented, deployed, and operationalized with appropriate safety and security measures in place.

According to our investigation, no universal measure or mitigation strategy exists that all cyber companies can apply to safeguard their digital environments against potential dangers. Organizations must be aware of the methods for reducing vulnerabilities, though. Additionally, it's important to give employees the right security training. It is anticipated that the findings will assist cyber businesses in gaining a deeper understanding of the vulnerabilities currently present in cyber security and the accompanying mitigation measures. This research provides academics and practitioners with a solid platform on which to build new cyber security solutions and address the aforementioned cyber security challenges.

## References

[1] R. P. Khandpur, T. Ji, S. Jan, G. Wang, C.-T. Lu, and N. Ramakrishnan, "Crowdsourcing Cybersecurity," *Proceedings of the 2017 ACM on Conference on Information and Knowledge Management*, Nov. 2017, **Published**, doi: 10.1145/3132847.3132866.

[2] Z. Li, D. Zou, S. Xu, H. Jin, H. Qi, and J. Hu, "VulPecker," *Proceedings of the 32nd Annual Conference on Computer Security Applications*, Dec. 2016, **Published**, doi: 10.1145/2991079.2991102.

[3] M. Cheng, M. Crow, and R. F. Erbacher, "Vulnerability analysis of a smart grid with monitoring and control system," *Proceedings of the Eighth Annual Cyber Security and Information Intelligence Research Workshop*, Jan. 2013, **Published**, doi: 10.1145/2459976.2460042.

[4] S. Zanero, "ULISSE, a network intrusion detection system," *Proceedings of the 4th annual workshop on Cyber security and information intelligence research: developing strategies to meet the cyber security and information intelligence challenges ahead*, May 2008, **Published**, doi: 10.1145/1413140.1413163.

[5] G. Werner, S. Yang, and K. McConky, "Time series forecasting of cyber attack intensity," *Proceedings of the 12th Annual Conference on Cyber and Information Security Research*, Apr. 2017, **Published**, doi: 10.1145/3064814.3064831.

[6] D. M. B. Masi, M. J. Fischer, J. F. Shortle, and C.-H. Chen, "Simulating network cyber attacks using splitting techniques," *Proceedings of the 2011 Winter Simulation Conference (WSC)*, Dec. 2011, **Published**, doi: 10.1109/wsc.2011.6148019.

[7] A. Okutan, S. J. Yang, and K. McConky, "Predicting cyber attacks with bayesian networks using unconventional signals," *Proceedings of the 12th Annual Conference on Cyber and Information Security Research*, Apr. 2017, **Published**, doi: 10.1145/3064814.3064823.

[8] A. Farraj, E. Hammad, and D. Kundur, "Impact of Cyber Attacks on Data Integrity in Transient Stability Control," *Proceedings of the 2nd Workshop on Cyber-Physical Security and Resilience in Smart Grids*, Apr. 2017, **Published**, doi: 10.1145/3055386.3055387.

[9] M. E. Kuhl, M. Sudit, J. Kistner, and K. Costantini, "Cyber attack modeling and simulation for network security analysis," *2007 Winter Simulation Conference*, Dec. 2007, **Published**, doi: 10.1109/wsc.2007.4419720.

[10] M. Gudo and K. Padayachee, "SpotMal," *Proceedings of the 2015 Annual Research Conference on South African Institute of Computer Scientists and Information Technologists*, Sep. 2015, **Published**, doi: 10.1145/2815782.2815812.

[11] Kim, Ikkyun, Daewon Kim, Byunggoo Kim, Yangseo Choi, Seongyong Yoon, Jintae Oh, and Jongsoo Jang. "A case study of unknown attack detection against Zero-day worm in the honeynet environment." In 2009 11th International Conference on Advanced Communication Technology, vol. 3, pp. 1715–1720. IEEE, 2009.

[12] F. Ahmadloo and F. R. Salmasi, "A cyber-attack on communication link in distributed systems and detection scheme based on H-infinity filtering," *2017 IEEE International Conference on Industrial Technology (ICIT)*, Mar. 2017, **Published**, doi: 10.1109/icit.2017.7915444.

[13] R. Aishwarya and S. Malliga, "Intrusion detection system- An efficient way to thwart against Dos/DDos attack in the cloud environment," *2014 International Conference on Recent Trends in Information Technology*, Apr. 2014, **Published**, doi: 10.1109/icrtit.2014.6996163.

[14] A. W. Al-Dabbagh, Y. Li, and T. Chen, "An Intrusion Detection System for Cyber Attacks in Wireless Networked Control Systems," *IEEE Transactions on Circuits and Systems II: Express Briefs*, vol. 65, no. 8, pp. 1049–1053, Aug. 2018, doi: 10.1109/tcsii.2017.2690843.

[15] M. Z. Alom and T. M. Taha, "Network intrusion detection for cyber security on neuromorphic computing system," *2017 International Joint Conference on Neural Networks (IJCNN)*, May 2017, **Published**, doi: 10.1109/ijcnn.2017.7966339.

[16] F. J. Aparicio-Navarro, K. G. Kyriakopoulos, Y. Gong, D. J. Parish, and J. A. Chambers, "Using Pattern-of-Life as Contextual Information for Anomaly-Based Intrusion Detection Systems," *IEEE Access*, vol. 5, pp. 22177–22193, 2017, doi: 10.1109/access.2017.2762162.

[17] P. Bhadre and D. Gothawal, "Detection and blocking of spammers using SPOT detection algorithm," *2014 First International Conference on Networks & Soft Computing (ICNSC2014)*, Aug. 2014, **Published**, doi: 10.1109/cnsc.2014.6906679.

[18] G. Bottazzi, E. Casalicchio, D. Cingolani, F. Marturana, and M. Piu, "MP-Shield: A Framework for Phishing Detection in Mobile Devices," *2015 IEEE International Conference on Computer and Information Technology; Ubiquitous Computing and Communications; Dependable, Autonomic and Secure Computing; Pervasive Intelligence and Computing*, Oct. 2015, **Published**, doi: 10.1109/cit/iucc/dasc/picom.2015.293.

[19] C.-M. Chen, H.-W. Hsiao, P.-Y. Yang, and Y.-H. Ou, "Defending malicious attacks in Cyber Physical Systems," *2013 IEEE 1st International Conference on Cyber-Physical Systems, Networks, and Applications (CPSNA)*, Aug. 2013, **Published**, doi: 10.1109/cpsna.2013.6614240.

[20] A. Chonka and J. Abawajy, "Detecting and Mitigating HX-DoS Attacks against Cloud Web Services," *2012 15th International Conference on Network-Based Information Systems*, Sep. 2012, **Published**, doi: 10.1109/nbis.2012.146.

[21] B. S. Kiruthika Devi, G. Preetha, G. Selvaram, and S. Mercy Shalinie, "An impact analysis: Real time DDoS attack detection and mitigation using machine learning," *2014 International Conference on Recent Trends in Information Technology*, Apr. 2014, **Published**, doi: 10.1109/icrtit.2014.6996133.

[22] M. Eslahi, H. Hashim, and N. M. Tahir, "An efficient false alarm reduction approach in HTTP-based botnet detection," *2013 IEEE Symposium on Computers & Informatics (ISCI)*, Apr. 2013, **Published**, doi: 10.1109/isci.2013.6612403.

[23] D. Gantsou, "On the use of security analytics for attack detection in vehicular ad hoc networks," *2015 International Conference on Cyber Security of Smart Cities, Industrial Control System and Communications (SSIC)*, Aug. 2015, **Published**, doi: 10.1109/ssic.2015.7245674.

[24] A. D. Hesar and M. A. Attari, "Simulating and analysis of cyber attacks on a BLPC network," *2014 Smart Grid Conference (SGC)*, Dec. 2014, **Published**, doi: 10.1109/sgc.2014.7150704.

[25] J. Hong, C.-C. Liu, and M. Govindarasu, "Integrated Anomaly Detection for Cyber Security of the Substations," *IEEE Transactions on Smart Grid*, vol. 5, no. 4, pp. 1643–1653, Jul. 2014, doi: 10.1109/tsg.2013.2294473.

[26] X. Hu *et al.*, "BAYWATCH: Robust Beaconing Detection to Identify Infected Hosts in Large-Scale Enterprise Networks," *2016 46th Annual IEEE/IFIP International Conference on Dependable Systems and Networks (DSN)*, Jun. 2016, **Published**, doi: 10.1109/dsn.2016.50.

[27] H. Ichise, Y. Jin, and K. Iida, "Analysis of via-resolver DNS TXT queries and detection possibility of botnet communications," *2015 IEEE Pacific Rim Conference on Communications, Computers and Signal Processing (PACRIM)*, Aug. 2015, **Published**, doi: 10.1109/pacrim.2015.7334837.

[28] I. Indre and C. Lemnaru, "Detection and prevention system against cyber attacks and botnet malware for information systems and Internet of Things," *2016 IEEE 12th International Conference on Intelligent Computer Communication and Processing (ICCP)*, Sep. 2016, **Published**, doi: 10.1109/iccp.2016.7737142.

[29] A. H. M. Jakaria, W. Yang, B. Rashidi, C. Fung, and M. A. Rahman, "VFence: A Defense against Distributed Denial of Service Attacks Using Network Function Virtualization," *2016 IEEE 40th Annual Computer Software and Applications Conference (COMPSAC)*, Jun. 2016, **Published**, doi: 10.1109/compsac.2016.219.

[30] G. Jin, F. Zhang, Y. Li, H. Zhang, and J. Qian, "A Hash-Based Path Identification Scheme for DDoS Attacks Defense," *2009 Ninth IEEE International Conference on Computer and Information Technology*, 2009, **Published**, doi: 10.1109/cit.2009.47.

[31] M. A. Khan, S. K. Pradhan, and H. Fatima, "Applying Data Mining techniques in Cyber Crimes," *2017 2nd International Conference on Anti-Cyber Crimes (ICACC)*, Mar. 2017, **Published**, doi: 10.1109/anti-cybercrime.2017.7905293.

[32] M. S. Khan, K. Ferens, and W. Kinsner, "A chaotic measure for cognitive machine classification of distributed denial of service attacks," *2014 IEEE 13th International Conference on Cognitive Informatics and Cognitive Computing*, Aug. 2014, **Published**, doi: 10.1109/icci-cc.2014.6921448.

[33] X. Kong, Y. Chen, H. Tian, T. Wang, Y. Cai, and X. Chen, "A Novel Botnet Detection Method Based on Preprocessing Data Packet by Graph Structure Clustering," *2016 International Conference on Cyber-Enabled Distributed Computing and Knowledge Discovery (CyberC)*, Oct. 2016, **Published**, doi: 10.1109/cyberc.2016.16.

[34] S. Misra, P. V. Krishna, H. Agarwal, A. Saxena, and M. S. Obaidat, "A Learning Automata Based Solution for Preventing Distributed Denial of Service in Internet of Things," *2011 International Conference on Internet of Things and 4th International Conference on Cyber, Physical and Social Computing*, Oct. 2011, **Published**, doi: 10.1109/ithings/cpscom.2011.84.

[35] F. Sanchez and Z. Duan, "A Sender-Centric Approach to Detecting Phishing Emails," *2012 International Conference on Cyber Security*, Dec. 2012, **Published**, doi: 10.1109/cybersecurity.2012.11.

[36] S. Shitharth and D. P. Winston, "A novel IDS technique to detect DDoS and sniffers in smart grid," *2016 World Conference on Futuristic Trends in Research and Innovation for Social Welfare (Startup Conclave)*, Feb. 2016, **Published**, doi: 10.1109/startup.2016.7583897.

[37] J.-H. Sun, T.-H. Jeng, C.-C. Chen, H.-C. Huang, and K.-S. Chou, "MD-Miner: Behavior-Based Tracking of Network Traffic for Malware-Control Domain Detection," *2017 IEEE Third International Conference on Big Data Computing Service and Applications (BigDataService)*, Apr. 2017, **Published**, doi: 10.1109/bigdataservice.2017.16.

[38] T. Velauthapillai, A. Harwood, and S. Karunasekera, "Global Detection of Flooding-Based DDoS Attacks Using a Cooperative Overlay Network," *2010 Fourth International Conference on Network and System Security*, Sep. 2010, **Published**, doi: 10.1109/nss.2010.68.

[39] C. Sun, J. Liu, X. Xu, and J. Ma, "A Privacy-Preserving Mutual Authentication Resisting DoS Attacks in VANETs," *IEEE Access*, vol. 5, pp. 24012–24022, 2017, doi: 10.1109/access.2017.2768499.

[40] L. Fan, Y. Wang, X. Cheng, and S. Jin, "Privacy Theft Malware Detection with Privacy Petri Net," *2012 13th International Conference on Parallel and Distributed Computing, Applications and Technologies*, Dec. 2012, **Published**, doi: 10.1109/pdcat.2012.113.

[41] H. Cui, Y. Zhou, C. Wang, Q. Li, and K. Ren, "Towards Privacy-Preserving Malware Detection Systems for Android," *2018 IEEE 24th International Conference on Parallel and Distributed Systems (ICPADS)*, Dec. 2018, **Published**, doi: 10.1109/padsw.2018.8644924.

[42] L. Xu, C. Jiang, N. He, Z. Han, and A. Benslimane, "Trust-Based Collaborative Privacy Management in Online Social Networks," *IEEE Transactions on Information Forensics and Security*, vol. 14, no. 1, pp. 48–60, Jan. 2019, doi: 10.1109/tifs.2018.2840488.

[43]T. Spyridopoulos, G. Karanikas, T. Tryfonas, and G. Oikonomou, “A game theoretic defence framework against DoS/DDoS cyber attacks,” *Computers & Security*, vol. 38, pp. 39–50, Oct. 2013, doi: 10.1016/j.cose.2013.03.014.
[44] S. Shitharth and D. P. Winston, “A Comparative Analysis between Two Countermeasure Techniques to Detect DDoS with Sniffers in a SCADA Network,” *Procedia Technology*, vol. 21, pp. 179–186, 2015, doi: 10.1016/j.protcy.2015.10.086.
[45] F. Wang, H. Wang, X. Wang, and J. Su, “A new multistage approach to detect subtle DDoS attacks,” *Mathematical and Computer Modelling*, vol. 55, no. 1–2, pp. 198–213, Jan. 2012, doi: 10.1016/j.mcm.2011.02.025.
[46] G. Varshney, M. Misra, and P. K. Atrey, “A phish detector using lightweight search features,” *Computers & Security*, vol. 62, pp. 213–228, Sep. 2016, doi: 10.1016/j.cose.2016.08.003.
[47] T. Liu *et al.*, “Abnormal traffic-indexed state estimation: A cyber–physical fusion approach for Smart Grid attack detection,” *Future Generation Computer Systems*, vol. 49, pp. 94–103, Aug. 2015, doi: 10.1016/j.future.2014.10.002.
[48] Y. Qiu, M. Ma, and S. Chen, “An anonymous authentication scheme for multi-domain machine-to-machine communication in cyber-physical systems,” *Computer Networks*, vol. 129, pp. 306–318, Dec. 2017, doi: 10.1016/j.comnet.2017.10.006.
[49] A. K. M.A. and J. C.D., “Automated multi-level malware detection system based on reconstructed semantic view of executables using machine learning techniques at VMM,” *Future Generation Computer Systems*, vol. 79, pp. 431–446, Feb. 2018, doi: 10.1016/j.future.2017.06.002.
[50] M. Noor, H. Abbas, and W. B. Shahid, “Countering cyber threats for industrial applications: An automated approach for malware evasion detection and analysis,” *Journal of Network and Computer Applications*, vol. 103, pp. 249–261, Feb. 2018, doi: 10.1016/j.jnca.2017.10.004